\documentclass{elsart}

\begin{document}

\newcommand{\nl}{\nonumber \\}
\newcommand{\be}{\begin{equation}}
\newcommand{\ee}{\end{equation}}
\newcommand{\bea}{\begin{eqnarray}}
\newcommand{\eea}{\end{eqnarray}}


\begin{frontmatter}
\title{  TARCER - A Mathematica program for the reduction of two-loop
  propagator integrals
}

\author{R. Mertig}\footnote{E-mail: $rolf@mertig.com$}
\address{Mertig Research \& Consulting, \\ 
Kruislaan 419, NL-1098 VA Amsterdam, \\ The Netherlands
}

\vspace{3.ex}

\author{R. Scharf}\footnote{E-mail: $scharf@itp.uni-leipzig.de$}
\address{
Institut f\"ur Theoretische Physik, Universit\"at Leipzig,\\
Augustusplatz 10/11, D-04109 Leipzig, \\ Germany
}

\vspace{3.ex}

\author{
\small
Submitted to Computer Physics Communications
\normalsize
}

\vspace{3.ex}

\begin{abstract}
TARCER is an implementation of the recurrence algorithm 
of O.V.~Tarasov for the reduction of two-loop propagator integrals 
with arbitrary masses to a small set of basis integrals.
The tensor integral reduction scheme is adapted to 
moment integrals emerging in operator matrix element calculations.
\end{abstract}

\end{frontmatter}

\newpage

\section{Program Summary}

\noindent
\emph{Title of program:} 
TARCER
\\ \\
\noindent
\emph{Version number:} 
1.0
\\ \\
\noindent
\emph{Available at} 
{\tt http://www.mertig.com/tarcer} \emph{or}\\
\phantom{\emph{Available at}} 
{\tt http://www.physik.uni-leipzig.de/TET/tarcer} \\
\\ 
\noindent
\emph{Programming Language:} 
{\it Mathematica} 3.0
\\ \\
\noindent
\emph{Platform:} 
Any platform supporting {\it Mathematica} 3.0
\\ \\
\noindent
\emph{Keywords:} 
Feynman integrals, two-loop corrections
\\ \\
\noindent
\emph{Nature of physical problem:} 
Reduction of Feynman integrals in pertubative two-loop calculations
\\ \\
\noindent
\emph{Method of solution:} 
The system of recurrence relations given by Tarasov \cite{tar}
\\ \\
\noindent
\emph{Limitations:} 
Rank of integrals 

\section{Introduction}

TARCER reduces two-loop propagator integrals with arbitrary masses 
to simpler basis integrals
using the reduction algorithm proposed 
by Tarasov \cite{tar,tar0}. 
The reduction of scalar products in the numerator of the integral
is extended to provide for the presence of an additional 
external vector that is lightlike.
For the reduction of scalar integrals
TARCER contains the complete set of recurrence relations given
in \cite{tar} and some additions for particular parameter configurations.
In some cases the number of basic integrals is reduced.
Further additions may easily be added. 

Due to a vast number of interrelations between the integrals considered 
it is not immediately obvious how to extract a set of recurrence relations 
that reduce the complexity of the integrals at each step such that 
one finally arrives at only a small set of basic 
integrals\footnote{
The fact that in principle all integrals considered 
must be expressible in terms of a \emph{finite} set of basic integrals 
may be seen as follows \cite{LaRe}:
Interpret the corresponding Feynman parameter integral representations 
as integrals over projective differential forms. 
All forms in question exist on the same differential manifold.
De Rham cohomology then implies that there exists
only a finite number of inequivalent forms on this manifold.
This in turn implies the above statement.}.
This was achieved by Tarasov.
The resulting relations are in some cases quite involved. 
Therefore  transcribing them into a program is an error prone process. 
But, whereas the extraction of adequate relations proves to be
quite difficult they can be checked in most cases much more
easily. For four or less propagators this may be done via a
Mellin-Barnes representation of the integral. In the present program
this is automatized in order to minimize input errors.

The emphasis of TARCER is not so much on the speed of evaluation.
Our main purpose was to use TARCER to check complicated 
moment integrals emerging from operator matrix element 
calculations for small individual moments ($m\leq5$). 
An example of this type of integral is given in section \ref{examples}. 
Conversely, those integrals, calculated independently by other means, 
served as a check for TARCER.

The massless case was checked using a FORM program \cite{hamberg}. 
The reader may also want to compare this program
with other existing fully analytic two-loop programs like
\cite{mincer}, \cite{shell2} and \cite{schorsch1,schorsch} 
or the original implementation in \cite{tar}.

\section{Notation}

TARCER reduces the following general type of integrals
to basic integrals.  The {\it Mathematica} notation, making all
dependencies explicit, is listed first:
\begin{eqnarray}
\lefteqn{
{\tt TFI[d, p^2, \Delta p, \{a, b\}, \{u, v, r, s, t\},
     \{ \{\nu_1,m_1\}, \{ \{\nu_2,m_2\}, \ldots, \{\nu_5,m_5\} \}] } 
= } 
\label{TFInotation1}
\nonumber \\
&& \frac{1}{\pi^d} \int\!\int
\frac{\ d^d k_1 d^dk_2 \ \ (\Delta k_1)^a \: (\Delta k_2)^b \:
 (k_1^2)^u \: (k_2^2)^v \: (p k_1)^r \: (p k_2)^s \: (k_1 k_2)^t}
{
[k_1^2-m_1^2]^{\nu_1}\ [k_2^2-m_2^2]^{\nu_2}\
[k_3^2-m_3^2]^{\nu_3}\ [k_4^2-m_4^2]^{\nu_4}\
[k_5^2-m_5^2]^{\nu_5}
} \; ,
\end{eqnarray}
with the abbreviations
$k_3 = k_1-p, k_4 = k_2-p$ and $k_5 = k_1 - k_2$.
The exponents $a, \ldots, t$ and the indices $\nu_1,\ldots,\nu_5$ 
are assumed to be nonnegative integers. 
$\Delta$ denotes a lightlike vector with $\Delta^2=0$.

If some of the subsets $\{a, b\}$ or $\{u, v, r, s, t\}$ of exponents
vanish we have the following reduced notations:
\begin{eqnarray}
\lefteqn{
{\tt TFI[d, p^2, \Delta p, \{0, 0\}, \{u, v, r, s, t\},
     \{ \{\nu_1,m_1\}, \{\nu_2,m_2\}, \ldots, \{\nu_5,m_5\} \}] }
= }
\nonumber \\
&&
{\tt TFI[d, p^2, \{u, v, r, s, t\},
     \{ \{\nu_1,m_1\}, \{\nu_2,m_2\}, \ldots, \{\nu_5,m_5\} \}] } \; ,
\label{TFInotation2}
\\ \nonumber \\
\lefteqn{
{\tt TFI[d, p^2, \Delta p, \{a, b\}, \{0, 0, 0, 0, 0\},
     \{ \{\nu_1,m_1\}, \{\nu_2,m_2\}, \ldots, \{\nu_5,m_5\} \}] } 
= }
\nonumber \\
&&
{\tt TFI[d, p^2, \Delta p, \{a, b\}, 
     \{ \{\nu_1,m_1\}, \{\nu_2,m_2\}, \ldots, \{\nu_5,m_5\} \}] } \; ,
\label{TFInotation3}
\\ \nonumber \\
\lefteqn{
{\tt TFI[d, p^2, \Delta p, \{0, 0\}, \{0, 0, 0, 0, 0\},
     \{ \{\nu_1,m_1\}, \{\nu_2,m_2\}, \ldots, \{\nu_5,m_5\} \}] } 
= }
\nonumber \\
&&
{\tt TFI[d, p^2,
     \{ \{\nu_1,m_1\}, \{\nu_2,m_2\}, \ldots, \{\nu_5,m_5\} \}] } \; .
\label{TFInotation4}
\end{eqnarray}
Furthermore, if a mass vanishes the argument $\{\nu_j,0\}$ 
of TFI may be replaced by the index $\nu_j$ alone.  
For example the fully massless integral with 
five proagators may be entered as 
${\tt TFI[d, p^2, \{1,1,1,1,1\}]}$.

In the course of evaluation all scalar products in the
numerator will be eliminated. 
If no more scalar products are present we keep
the notation in line with \cite{tar}: 
\begin{eqnarray}
\lefteqn{
{\tt TFI[d, p^2, \{ \{\nu_1,m_1\}, \{\nu_2,m_2\}, \ldots, 
\{\nu_5,m_5\} \}] }  
= } 
\nonumber \\
&&
F^{(d)}_{\nu_1 \nu_2 \nu_3 \nu_4 \nu_5} 
= 
\frac{1}{\pi^d}
\int\!\int
\frac{d^d k_1 d^dk_2}{
[k_1^2-m_1^2]^{\nu_1}\ [k_2^2-m_2^2]^{\nu_2}\
\cdots
[k_5^2-m_5^2]^{\nu_5}
} \; ,
\\ \nonumber \\
\lefteqn{
{\tt TVI[d, p^2, \{ \{\nu_1,m_1\},  \{\nu_2,m_2\}, 
\{\nu_3,m_3\}, \{\nu_4,m_4\} \}] }  
= }
\nonumber \\
&&
V^{(d)}_{\nu_1 \nu_2 \nu_3 \nu_4 } 
=
\frac{1}{\pi^d}
\int\!\int
\frac{d^d k_1 d^dk_2}{
[k_5^2-m_1^2]^{\nu_1}\ [k_2^2-m_2^2]^{\nu_2}\
[k_3^2-m_3^2]^{\nu_3}\ [k_4^2-m_4^2]^{\nu_4}
} \; ,
\\ \nonumber \\
\lefteqn{
{\tt TJI[d, p^2, \{ \{\nu_1,m_1\}, \{\nu_2,m_2\}, \{\nu_3,m_3\} \}] }  
= }
\nonumber \\
&&
J^{(d)}_{\nu_1 \nu_2 \nu_3 } 
=
\frac{1}{\pi^d}
\int\!\int
\frac{d^d k_1 d^dk_2}{
[k_1^2-m_1^2]^{\nu_1}\ [k_5^2-m_2^2]^{\nu_2}\ [k_4^2-m_3^2]^{\nu_3}
} \; ,
\\ \nonumber \\
\lefteqn{
{\tt TJI[d, 0, \{ \{\nu_1,m_1\}, \{\nu_2,m_2\}, \{\nu_3,m_3\} \}] }  
= }
\nonumber \\
&&
K^{(d)}_{\nu_1 \nu_2 \nu_3 } 
=
\frac{1}{\pi^d}
\int\!\int
\frac{d^d k_1 d^dk_2}{
[k_1^2-m_1^2]^{\nu_1}\ [k_5^2-m_2^2]^{\nu_2}\ [k_2^2-m_3^2]^{\nu_3}
}
\end{eqnarray}
and
\begin{eqnarray}
\lefteqn{
{\tt TBI[d, p^2, \{ \{\nu_1,m_1\}, \{\nu_2,m_2\} \}] }  
= } 
\nonumber \\
&& 
B^{(d)}_{\nu_1 \nu_2} =
\frac{1}{\pi^{d/2}}
\int\
\frac{d^d k_1}{[k_1^2-m_1^2]^{\nu_1}\ [k_3^2-m_2^2]^{\nu_2}} \; ,
\\ \nonumber \\
\lefteqn{
{\tt TAI[d, 0, \{ \{\nu_1,m_1\} \}] }  
= } 
\nonumber \\ 
&&
A^{(d)}_{\nu_1} =
\frac{1}{\pi^{d/2}}
\int\
\frac{d^d k_1}{[k_1^2-m_1^2]^{\nu_1}} \; .
\end{eqnarray}
Input for TARCER should be prepared in terms of the {\tt TFI}-notation.

\section{Elimination of numerators}

The TARCER-function {\tt TarcerRecurse} is the principal function
which, when applied to an expression involving the above integrals,
performs the complete reduction to the set of basic integrals.
In the first step the numerator of the integrand in the
{\tt TFI}-integrals is simplified as far as possible 
by standard manipulations until an irreducible numerator of the
form $(\Delta k_1)^a \: (\Delta k_2)^b \: (p k_1)^r \: (p k_2)^s$
results. The remaining integrals are of the form
\begin{equation}
I^{(d)}_{a b r s} =
{\tt TFI[d, p^2, \Delta p, \{a, b\}, \{0, 0, r, s, 0\},
     \{ \{\nu_1,m_1\}, \ldots, \{\nu_5,m_5\} \}] } \;.
\label{irreducible}
\end{equation}
The critical observation of Tarasov is that integrals containing
irreducible numerators may be rewritten in terms of scalar integrals 
in higher space-time dimensions and that those can later be reduced again
to scalar integrals in the original space-time dimension. 
Integrals of the form (\ref{irreducible}) we rewrite as 
\begin{equation}
I^{(d)}_{a b r s}
=
T_{a b r s} (p^2, \Delta p, \{\partial\}, {\bf d^+}) \;
I^{(d)}_{0 0 0 0} \;,
\label{Trelation}
\end{equation}
by employing an operator $T$ that is a polynomial in  
the operator ${\bf d^+}$ representing 
a shift $d \rightarrow d+2$ in dimension
and in the mass derivatives
$\partial_j = \partial / \partial m_j^2$.
With the lightlike vector $\Delta$ present, 
the required $T$-operator is given by 
a generalization of eq.~(25) of
\cite{tar}\footnote{We have 
$\rho=- {\bf d^+}$ instead of $-1/\pi^2 \, {\bf d^+}$
due to the factor $1/\pi^d$ in the normalization
of the integrals in (\ref{Trelation}).
}:
\begin{eqnarray}
\lefteqn{
T_{a b r s} (p^2, \Delta p, \{\partial\}, {\bf d^+}) =
    \left( \frac{\partial}{ i \partial \gamma_1 } \right )^a
    \left( \frac{\partial}{ i \partial \gamma_2 } \right )^b
    \left( \frac{\partial}{ i \partial \beta_1 } \right )^r
    \left( \frac{\partial}{ i \partial \beta_2 } \right )^s  
}
 \nonumber \\ 
 \nonumber \\ 
&& 
    \times 
\exp 
      \left[ i 
               \left\{ (\beta_1 p^2 + \gamma_1 (\Delta p) ) \: Q_1 
                   + (\beta_2 p^2 + \gamma_2 (\Delta p) ) \: Q_2  
\right. \right.
\nonumber \\ 
&& 
\phantom{\times \exp i} 
\left. \left.
                   +  \beta_1 ( \beta_1 p^2 + 2 \gamma_1 (\Delta p) 
                              ) \: Q_{11}
                   +  \beta_2 ( \beta_2 p^2 + 2 \gamma_2 (\Delta p) 
                              ) \: Q_{22}
\right. \right.
\nonumber \\ 
&& 
\phantom{\times \exp i}
\left. \left.
                   +  ( \beta_1 \beta_2 p^2 + 
                        (\beta_1 \gamma_2 + \beta_2 \gamma_1) (\Delta p) 
                              ) \: Q_{12} 
          \right\} \rho
      \right] 
 \left|_{{\beta_i=0, \: \gamma_i=0}
 \atop {\rho=- {\bf d^+}} }\right. \; ,
\label{Toperator}
\end{eqnarray}
where the $Q$'s are polynomials in $\partial_j$
\begin{eqnarray}
4 i \: Q_{11}=\partial_2+\partial_4+\partial_5 \; ,
&&
- Q_1 = \partial_3\partial_5+\partial_4\partial_5+\partial_2\partial_3
+\partial_3\partial_4 \; , 
\nonumber \\
4 i \: Q_{22} = \partial_1+\partial_3+\partial_5 \; ,
&&
- Q_2 = \partial_4\partial_5+\partial_3\partial_5+\partial_1\partial_4
+\partial_3\partial_4 \; ,
\nonumber \\
2 i \: Q_{12} = \partial_5 \; .
&&
\end{eqnarray}
 
\section{Recurrence relations}

Once all irreducible numerators are eliminated the next
step taken by the function {\tt TarcerRecurse} is to repeatedly
apply the recurrence relations that reduce the exponents of
the scalar propagators in the integrals until no further
reduction is possible. All recurrence relations explicitly
or implicitly given by Tarasov are implemented.

For certain classes of on-shell integrals we found it
necessary to supplement the recurrence relations
by some additions in order to achieve maximal reduction.
Those additional relations are listed in the following.
The relations obtained by permutations of indices and 
masses are also implemented.
The operators ${\bf 1^\pm}, {\bf 2^\pm}, \ldots$ act on an integral 
by in-/decreasing the first, second, etc.  index by one unit. 

For  $m_1=0$
and $\nu_2>1$:
\bea
&&
2\ m_{2}^{2}\ (1-{{\nu }_2})\ {{\nu }_2}\
          {\bf 2^+}J_{{{\nu }_1}{{\nu }_2}{{\nu }_3}}^{(d)}
=
\nl
&&~~~~~~~
(d-2\ {{\nu }_2})\ (1-{{\nu }_2})\ J_{{{\nu }_1}{{\nu }_2}{{\nu }_3}}^{(d)}-
{{\nu }_1}\ (-d+2\ {{\nu }_1}+2)\
 {\bf 1^+}{\bf 2^-}J_{{{\nu }_1}{{\nu }_2}{{\nu }_3}}^{(d)} \;.
\label{req100}
\eea
For $p^2 = m_1^2$, $m_2 = m_3 = 0$
and $\nu_1>0$:
\bea
&&
\nl
&&{{\nu }_1}\ (d-2\ {{\nu }_2}-2\ {{\nu }_3})\ (d-{{\nu }_2}-
{{\nu }_3}-1)\ {\bf 1^+}
 J_{{{\nu }_1}{{\nu }_2}{{\nu }_3}}^{(d)} 
=
\nl
&&~~~~~~~
-{{\nu }_2}\ (-d+2\ {{\nu }_2}+2)\ (-2\ d+{{\nu }_1}+2\ {{\nu }_2}+
2\ {{\nu }_3}+2){\bf 2^+}J_{{{\nu }_1}{{\nu }_2}{{\nu }_3}}^{(d)} 
\; .
\label{req99}
\eea
For $m_1=m_2=0$, $p^2 = m_3^2$ 
and $\nu_1>1$:
\bea
&&
J_{{{\nu }_1}{{\nu }_2}{{\nu }_3}}^{(d)} = 
-\frac{
  \left( d - \nu_{1} - \nu_{2} \right) \,
   \left( -2 - d + 2\,\nu_{1} + 2\,\nu_{2} \right)
}{
2\,{{m_{3}}^2}\,\left( d - 2\,\nu_{1} \right) \,
   \left( -1 + \nu_{1} \right) 
}
\nl
&&
~~~~\times \,
\frac{
   \left( 3\,d - 2\,\nu_{1} - 2\,\nu_{2} - 2\,\nu_{3} \right) \,
   \left( -1 - d + \nu_{1} + \nu_{2} + \nu_{3} \right) 
}{ \left( 2\,d - 2\,\nu_{1} - 2\,\nu_{2} - \nu_{3} \right) \,
   \left( 1 + 2\,d - 2\,\nu_{1} - 2\,\nu_{2} - \nu_{3} \right) 
} \; {\bf 1^-}
J_{{{\nu }_1}{{\nu }_2}{{\nu }_3}}^{(d)} 
\;.
\label{req201}
\eea
For $m_1=0$, $m_2=m_3$, $p^2 = 0$ and $\nu_1>0$:
\bea
K_{{{\nu }_1}{{\nu }_2}{{\nu }_3}}^{(d)} =
&&
\frac{\left( d - 2\,\left( -1 + \nu_{1} + \nu_{2} \right)  \right) \,
      \left( 1 + d - \nu_{1} - \nu_{2} - \nu_{3} \right) 
     }{2\,{{m_{2}}^2}\,
      \left( d - 2\,\nu_{1} \right) \,
      \left( 1 + d - 2\,\nu_{1} - \nu_{2} - \nu_{3} \right) \,
      }
\nl
&&~~~~\times \,
\frac{
      \left( d - 2\,\left( -1 + \nu_{1} + \nu_{3} \right)  \right) 
     }{
      \left( 2 + d - 2\,\nu_{1} - \nu_{2} - \nu_{3} \right) 
      }  \; {\bf 1^- } \; K_{{{\nu }_1}{{\nu }_2}{{\nu }_3}}^{(d)} 
\; .
\label{req203}
\eea
For $m_1=m_2=m_3$, $p^2 = m_1^2$
and $\nu_1>0$:
\bea
&&
16\ m_{1}^{2}\ {{\nu }_1}\ (d-{{\nu }_1}-{{\nu }_2}-{{\nu }_3})\
{\bf 1^+}J_{{{\nu }_1}{{\nu }_2}{{\nu }_3}}^{(d)} =
\nl
&&~~~~~~~
-(1+3\ d-3\ {{\nu }_1}-4\ {{\nu }_2})\ {{\nu }_3}\
{\bf 1^-}{\bf 3^+}J_{{{\nu }_1}{{\nu }_2}{{\nu }_3}}^{(d)}
\nl
&&~~~~~~~
-
{{\nu }_2}\ (1+3\ d-3\ {{\nu }_1}-4\ {{\nu }_3})\
{\bf 1^-}{\bf 2^+}J_{{{\nu }_1}{{\nu }_2}{{\nu }_3}}^{(d)}
\nl
&&~~~~~~~
+(-1+2\ d-{{\nu }_1}-2\ {{\nu }_2})\ {{\nu }_3}\
{\bf 2^-} {\bf 3^+} J_{{{\nu }_1}{{\nu }_2}{{\nu }_3}}^{(d)}
\nl
&&~~~~~~~
-
\left(-6\ {d^2}+22\ d\ {{\nu }_1}-16\ \nu _{1}^{2}-
{{\nu }_2}+
6\ d\ {{\nu }_2}-13\ {{\nu }_1}\ {{\nu }_2}
\right.
\nl
&&~~~~~~~
-{{\nu }_3}+
6\ d\ {{\nu }_3}-13\ {{\nu }_1}\ {{\nu }_3}
-4\ {{\nu }_2}\ {{\nu }_3})\ J_{{{\nu }_1}{{\nu }_2}{{\nu }_3}}^{(d)}
\nl
&&~~~~~~~
+
{{\nu }_2}\ (-1+2\ d-{{\nu }_1}-2\ {{\nu }_3})\
{\bf 2^+}{\bf 3^-}\, J_{{{\nu }_1}{{\nu }_2}{{\nu }_3}}^{(d)}
\; .
\label{req200}
\eea
Note that in the last case all
$J_{{{\nu }_1}{{\nu }_2}{{\nu }_3}}^{(d)}$
are eventually reduced to
$J_{1 1 1}^{(d)}$ and $(A_1^{(d)})^2$.
That is, the number of basic two-loop integrals
is smaller than in the general case.

Other special configurations may require still further 
additional relations.

\section{Verification of recurrence relations}
As stated in \cite{tar} the derivation `of these
relations is rather tedious and for brevity of the presentation 
will be omitted'. Therefore, we rederived most of the recurrence relations.
Furthermore, an automatic verification of recurrence relations 
for the integrals of type $J$ or $V$ with three
and four propagators in the general mass case 
is possible via the kernel of the
respective Mellin-Barnes type integral-representations.

By replacing the massive scalar propagators in the momentum space
integral with their respective Mellin-Barnes representation and 
interchanging integrations, 
as already described in \cite{mellin} for the case $\nu_i=1$, 
we obtain the Mellin-Barnes representation of $J$
for general $\nu_i >0$: 
\begin{equation}
J_{\nu_1 \nu_2 \nu_3}^{(d)} = 
\frac{1}{(2 \pi i)^{3}}
\int\!\!\!\!\int\limits_{-i\infty}^{+i\infty}\!\!\!\!\int
\psi_{\nu_1 \nu_2 \nu_3}^{(d)}(s_1,s_2,s_3) \,
(-p^2)^{d-\nu-s} 
\prod_{i=1}^{3} \,
(m_i^2)^{s_i} \; d s_i \;,
\label{Jmellin}
\end{equation}
with the kernel
\begin{eqnarray}
\lefteqn{
\psi_{\nu_1 \nu_2 \nu_3}^{(d)}(s_1,s_2,s_3) = 
} \\
&&
(-1)^{1+\nu} 
\frac{\Gamma(\nu+s-d)}{\Gamma(3 d/2-\nu-s)}
\prod_{i=1}^{3} 
\frac{\Gamma(-s_i) \Gamma(d/2-\nu_i-s_i)}{\Gamma(\nu_i)} \;\; ,
\nonumber
\end{eqnarray}
where $\nu = \nu_1 + \nu_2 + \nu_3$ and
$s = s_1 + s_2 + s_3$. 
As usual the integration contours separate the poles of the
gamma functions in the left and right half planes.

The Mellin-Barnes representation of $V$ 
contains four $s$-integrations in an analogous manner
with the kernel
\begin{eqnarray}
\lefteqn{
\psi_{\nu_1 \nu_2 \nu_3 \nu_4}^{(d)}(s_1,s_2,s_3,s_4) =
} \\
&&
(-1)^{1+\nu} 
\frac{
\Gamma(\nu_1+\nu_3+s_1+s_3-d/2)
\Gamma(d-\nu_1-\nu_3-\nu_4-s_1-s_3-s_4)
}
{
\Gamma(\nu_1+\nu_3+\nu_4+s_1+s_3+s_4-d/2)
\Gamma(d-\nu_1-\nu_3-s_1-s_3)
}
\nonumber
\\
&&
\phantom{(-1)^{1}}
\times
\frac{\Gamma(\nu+s-d)}{\Gamma(3 d/2-\nu-s)}
\frac{\Gamma(-s_4) \Gamma(\nu_4+s_4)}{\Gamma(\nu_4)}
\prod_{i=1}^{3} 
\frac{\Gamma(-s_i) \Gamma(d/2-\nu_i-s_i)}{\Gamma(\nu_i)} 
\; ,
\nonumber
\end{eqnarray}
where $\nu = \nu_1 + \nu_2 + \nu_3 + \nu_4$ and
$s = s_1 + s_2 + s_3 + s_4$.

Insertion of the above integral representation for $J$
into a vanishing combination of $J$'s 
as provided by a conjectured recurrence relation
yields under the integral a sum of the form
\begin{equation}
\sum_{n_1, n_2, n_3} 
\phi_{n_1 n_2 n_3}(s_1,s_2,s_3) \,
(-p^2)^{-n-s} 
(m_1^2)^{s_1+n_1} (m_2^2)^{s_2+n_2} (m_3^2)^{s_3+n_3} \;,
\end{equation}
where the $n_i$'s run over
a finite set of integers and $n = n_1+n_2+n_3$. 
Here each $\phi$ represents a sum over $\psi$'s with 
various values of the indices $\nu_i$ and $d$.  
By an appropriate shift of integration contours for each 
term separately accompanied by a change of variables 
$s_i \rightarrow s_i-n_i$ one extracts a global factor:
\begin{equation}
(-p^2)^{-s} (m_1^2)^{s_1} (m_2^2)^{s_2} (m_3^2)^{s_3} 
\sum_{n_1, n_2, n_3} 
\phi_{n_1 n_2 n_3} (s_1-n_1,s_2-n_2,s_3-n_3)  \;.
\label{integrand}
\end{equation}
The sum in (\ref{integrand}) uniquely determines the coefficients
of the large $p^2$ expansion of the Mellin-Barnes integral
over (\ref{integrand}) in terms of 
(fractional) powers of $m_i^2/(-p^2)$.
Therefore, if the recurrence relation
is indeed correct and this 
integral vanishes then the above sum
has to vanish identically.

This observation provides a method to verify
recurrence relations for both $J$ and $V$, 
where a similar Ansatz applies. The method is
suitable for recurrence relations that are 
valid for general values of $m_i^2$ and $p^2$. 
It can also be adapted to the case when some of 
the $m_i^2$ or $p^2$ are zero. 
It is, however, not applicable to relations that hold only for
particular parameter configurations, e.g., when some 
kinematical determinants vanish. 

This verification-procedure is implemented in TARCER 
through the functions {\tt CheckTJIRecursion} and {\tt CheckTVIRecursion}.
The notebook {\tt TARCER.nb} also contains comments for each particular
recursion as to whether this check is applicable or not.

Despite the complexity of the equations presented in \cite{tar} 
we found only one non-obvious\footnote{
The lower limit of the sum in (46) should read $j=1$ and
$m_3$ in (66) needs to be replaced by $m_3^2$.}
misprint: In eq.~(67) the global sign of the right hand 
side has to be reversed.

\section{Usage and examples}
\label{examples}

The {\it Mathematica} 3.0 notebook {\tt TARCER.nb} 
contains the complete source code.
Upon evaluation the recurrence relations are 
incorporated into the function {\tt TarcerRecurse}.
All $T$-operators (\ref{Toperator}) up to
$\{a+b,r+s\} \leq \mbox{{\tt \$RankLimit}}$
are constructed and the corresponding relations 
(\ref{Trelation}) explicitly generated.
The value of {\tt \$RankLimit} 
can be set in the prologue section.

The evaluation of {\tt TARCER.nb} generates
a binary file {\tt tarcer.mx} 
containing all functions and definitions 
in internal {\it Mathematica} format.
An explicit integral reduction only requires 
the latter to be loaded directly into
{\it Mathematica} 3.0.
A set of preproduced tarcer*.mx 
files for different operating systems is also available.
Therefore one only needs to run the {\tt TARCER.nb} 
notebook if one wants to increase {\tt \$RankLimit}
or wishes to modify the program.

{\tt TarcerRecurse} is the main function that performs the integral
reduction. 
Its usage is simply as follows: It can be 
applied to any expression involving 
the functions {\tt TFI, TVI, TJI, TBI} and 
{\tt TAI} whose first argument is a symbol {\tt d}.

For a simple example, consider the following on-shell integral 
where $p^2=M^2$:
\begin{equation}
\frac{1}{\pi^d} \int\!\int
\frac{d^d k_1 d^dk_2}
{
k_1^2\ [k_2^2-M^2]\ [(k_1-p)^2-M^2]\ (k_2-p)^2\ [(k_1-k_2)^2-M^2]
} \; .
\end{equation}
Application of {\tt TarcerRecurse} 
to the corresponding input form (\ref{TFInotation4})
yields:
\begin{eqnarray}
\lefteqn{ 
{\tt
TarcerRecurse[
TFI[d, M^2, \{ 1, \{1,M\}, \{1,M\}, 1, \{1,M\}\}]
] 
}
}
\nonumber \\ &&
\nonumber \\ &&
-\frac{
3\ (d - 2)^2\ (5\ d-18)\
 \left({\bf A}_{\{1,M\}}^{(d)}\right)^2
}{32\  (d - 4)^2\ (d - 3)\ M^6} \;
+ 
\nonumber
\\ &&
\frac{
 (3\ d - 10)\ (3\ d - 8)\ 
 {\bf J}_{\{1,M\}, \{1,0\}, \{1,0\}}^{(d)}
}{8\ (d - 4)\ (2\ d - 7)\ M^4} 
+
\nonumber
\\ &&
\frac{(3 d-10)\ (3 d-8 )\ 
{\bf J}_{\{1,M\}, \{1,M\}, \{1,M\}}^{(d)}
}{16\ (d - 4)^2\ M^4}
\end{eqnarray}
Note that the display format of TARCER explicitly shows
the masses together with the indices.

The function {\tt TarcerExpand} inserts explicit results
for some basis integrals as specified by the option 
{\tt TarcerReduce}.
A second argument to  {\tt TarcerExpand} must be given in form of 
a rule, like ${\tt d \rightarrow 4 + \varepsilon}$. 
Then an expansion of the first argument of {\tt TarcerExpand} in 
the sole variable specified, here $\varepsilon$, around $0$ will be performed. 

Applying {\tt TarcerExpand} to {\tt \%}, i.e. to the previous expression,
yields:
\begin{eqnarray}
\lefteqn{
{\tt TarcerExpand[\mbox{{\tt \%}}, d \rightarrow 4 + \varepsilon] }
}
\nonumber \\ &&
\nonumber \\ &&
(M^2)^{\varepsilon-1} 
S_{\varepsilon}^2
\left(6\ \zeta(2)\, \log(2) - \frac{3\ \zeta(3)}{2} \right)
\end{eqnarray}
Here $S_{\varepsilon} = e^{\gamma_{E}\ (d-4)/2}$ is 
used.\footnote{$\gamma_{E}$ is the Euler-constant} 
This result is of course well known, see \cite{broad}
where the expansion up to $\mathcal{O}(\varepsilon^3)$ is given.

~\\
As a somewhat more demanding example consider 
another on-shell integral with $p^2=M^2$ containing $\Delta k_1$
in the numerator:
\begin{equation}
\frac{1}{\pi^d} \int\!\int
\frac{\ d^d k_1 d^dk_2 \ \ (\Delta k_1)^m}
{
[k_1^2-M^2]^2\ k_2^2\ (k_1-p)^2\ [(k_2-p)^2-M^2]\ [(k_1-k_2)^2-M^2]
} \; .
\end{equation}
For general $m$ one can, with some effort, express this integral
in terms of a moment integral, Laurent-expanded around $d=4$: 
\begin{eqnarray}
\lefteqn{
(M^2)^{d-6}\ (\Delta p)^{m}\
S_{\varepsilon}^2\
\int_0^1 dx\ x^{m} 
\bigg(
\bigg(
\frac{1}{2}\ +\frac{3}{8}\ \zeta (2)\ -
\frac{1}{2\ {{(d-4)}^2}}
\bigg)\ \delta (1-x)\ + 
}
\nonumber
\\ &&
\hspace{1.em}
\bigg(
-\zeta (2)
+\frac{\log(1-x)}{2\ (1-x)}
-\frac{1}{4}\ {{\log}^2}(1-x)
-\frac{\log(x)}{2\ (1+x)}
-\frac{\log(x)}{2\ {{(1-x)}^2}}
+
\nonumber
\\ && \hspace{2.em}
\bigg( 1 - \frac{1}{{{(1-x)}^2}} \bigg) 
\bigg(
\frac{1}{2}\ \log(x)\ \log(1-x) + \frac{3}{4}\ {{\log}^2}(x) 
\bigg) 
-
\nonumber
\\ && \hspace{2.em}
\bigg( 1 - \frac{2}{{{(1-x)}^2}} \bigg) 
\bigg(
\frac{1}{2}\ \zeta (2) 
+ \log(x)\ \log(1+x) + {{\mbox{Li}}_2}(-x) \bigg)
+
\nonumber
\\ && \hspace{2.em}
\bigg( 1 - \frac{1}{2\ {{(1-x)}^2}} \bigg) \
{{\mbox{Li}}_2}(1-x)
\bigg) 
+ \mathcal{O} (d-4)
\bigg) \ .
\end{eqnarray}
TARCER was originally written to check this kind
of general expressions for individual moments.
Consider for example the moment $m=1$ and apply
{\tt TarcerRecurse} to the input form (\ref{TFInotation3}):
\begin{eqnarray}
\lefteqn{ 
{\tt
TarcerRecurse[
TFI[d, M^2, \Delta p, \{1, 0\}, 
   \{ \{2,M\}, 1, 1, \{1,M\}, \{1,M\}\}]
] 
}
}
\nonumber \\ &&
\nonumber \\ &&
-\frac{
(d - 2)^2\ (27\ d^4 - 404\ d^3 + 2175\ d^2 - 4902\ d + 3776)\
\Delta p \ \left({\bf A}_{\{1,M\}}^{(d)}\right)^2
}{256\ (d - 5)^2\ (d - 4)^2\ (d - 3)\ M^8}
-
\nonumber
\\ &&
\frac{
(d - 2)\ (3\ d - 10)\ (3\ d - 8) \
\Delta p \  {\bf J}_{\{1,M\}, \{1,0\}, \{1,0\}}^{(d)}
}{32\ (d - 5)\ (d - 4)\ (2 d - 7)\ M^6} 
+
\nonumber
\\ &&
\frac{d\ (3\ d - 10)\ (3\ d - 8)\ 
\Delta p \; {\bf J}_{\{1,M\}, \{1,M\}, \{1,M\}}^{(d)}
}{128\ (d - 4)^2\ M^6}
\end{eqnarray}
Expansion with {\tt TarcerExpand} yields:
\begin{eqnarray}
\lefteqn{
{\tt TarcerExpand[\mbox{{\tt \%}}, d \rightarrow 4 + \varepsilon] }
}
\nonumber \\ &&
\nonumber \\ &&
(M^2)^{\varepsilon-2} \Delta p \; S_{\varepsilon}^2\left(3\, \log(2)\ \zeta(2) -
\frac{11\ \zeta(2)}{8}-\frac{1}{2\ \varepsilon^2}  - \frac{3\ \zeta(3)}{4} +\frac{1}{2}
\right)
\end{eqnarray}
Which agrees with the analytic integration 
of the moment-integral above for $m=1$.

~\\
For a massive off-shell example consider the integral
\begin{equation}
\frac{1}{\pi^d} \int\!\int
\frac{\ d^d k_1 d^dk_2 \ \ (k_1^2)^2}
{
[k_2^2-m_4^2]\ [(k_1-p)^2-m_2^2]\ [(k_1-k_2)^2-m_3^2]
} \; ,
\end{equation}
denoted $Y_{234}^{11}$ in \cite{schorsch}.
Acting with
{\tt TarcerRecurse} on the input form (\ref{TFInotation2})
yields:
\begin{eqnarray}
\lefteqn{ 
{\tt
TarcerRecurse[
TFI[d, p^2, \{2, 0, 0, 0, 0\}, 
   \{ 0, \{1, m_4\}, \{1, m_2\}, 0, \{1, m_3\}\}]
] 
}
}
\nonumber \\ &&
\nonumber \\ &&
\frac{
(d\ ({p^2}+m_{2}^{2}+m_{3}^{2}+7\ m_{4}^{2})-
12\ m_{4}^{2})\ 
{\bf A}_{\{1,{m_2}\}}^{(d)}\ 
{\bf A}_{\{1,{m_3}\}}^{(d)}
}{3\ (3\ d-4)}+  
\nonumber \\ &&
\frac{
(d\ ({p^2}+m_{2}^{2}+7\ m_{3}^{2}+m_{4}^{2})-
12\ m_{3}^{2})\ 
{\bf A}_{\{1,{m_2}\}}^{(d)}\ 
{\bf A}_{\{1,{m_4}\}}^{(d)}
}{3\ (3\ d-4)}+  
\nonumber \\ &&
\frac{
((7\ d-12)\ m_{2}^{2}+
d\ ({p^2}+m_{3}^{2}+m_{4}^{2}))\ 
{\bf A}_{\{1,{m_3}\}}^{(d)}\ 
{\bf A}_{\{1,{m_4}\}}^{(d)}
}{3\ (3\ d-4)}+  
\nonumber \\ &&
\frac{1}{
3\ (d-2)\ (3\ d-4)}
\big(
(
(d-2)\ d\ {p^4}+
2\ (d-4)\ (d-3)\ (m_{3}^{2}+m_{4}^{2})\ {p^2}+  
\nonumber \\ &&
\hspace{2.em} 
(d\ (25\ d-102)+96)\ m_{2}^{4}-
2\ (d-3)\ 
(
d\ m_{3}^{4}+
\nonumber \\ &&
\hspace{2.em} 
2\ (7\ d-12)\ m_{4}^{2}\ m_{3}^{2}+
d\ m_{4}^{4}
)+  
2\ m_{2}^{2}\ 
(
(19\ (d-4)\ d+72)\ {p^2}-
\nonumber \\ &&
\hspace{2.em} 
(d-3)\ (5\ d-12)\ (m_{3}^{2}+m_{4}^{2})
)
)\ 
\:  
{\bf J}_{\{1,{m_4}\}\{1,{m_3}\}\{1,{m_2}\}}^{(d)}
\big)+  
\nonumber \\ &&
\frac{1}{3\ (d-2)\ (3\ d-4)}
\big(
2\ m_{3}^{2}\ 
(
-3\ (3\ d-4)\ m_{2}^{4}+
2\ ((18-11\ d)\ {p^2}+
\nonumber \\ &&
\hspace{2.em} 
2\ (d-3)\ m_{3}^{2})\ m_{2}^{2}-  
12\ m_{4}^{2}\ 
( -{p^2}+ 3\ m_{3}^{2}+ m_{4}^{2}
)+
\nonumber \\ &&
\hspace{2.em} 
d\ 
( 
-{p^4}-
4\ m_{4}^{2}\ {p^2}+
m_{3}^{4}+
9\ m_{4}^{4}+
22\ m_{3}^{2}\ m_{4}^{2}
)
)\   
\;
{\bf J}_{\{2,{m_3}\}\{1,{m_4}\}\{1,{m_2}\}}^{(d)}
\big)+  
\nonumber \\ &&
\frac{1}{3\ (d-2)\ (3\ d-4)}
\big(
2\ m_{4}^{2}\ 
(
-3\ (3\ d-4)\ m_{2}^{4}+
2\ ((18-11\ d)\ {p^2}+
\nonumber \\ &&
\hspace{2.em} 
2\ (d-3)\ m_{4}^{2})\ m_{2}^{2}+  
3\ (3\ d-4)\ m_{3}^{4}+
2\ m_{3}^{2}\ ((11\ d-18)\ m_{4}^{2}-
\nonumber \\ &&
\hspace{2.em} 
2\ (d-3)\ {p^2})+
d\ (m_{4}^{4}-{p^4})
)\   
\:
{\bf J}_{\{2,{m_4}\}\{1,{m_3}\}\{1,{m_2}\}}^{(d)}\big) -
\nonumber \\ &&
\frac{1}{3\ (d-2)\ (3\ d-4)}
\big(
8\ m_{2}^{2}\ (m_{2}^{2}-{p^2})\   
(
(2\ d-3)\ {p^2}+
(2\ d-3)\ m_{2}^{2}-
\nonumber \\ &&
\hspace{2.em} 
(d-3)\ (m_{3}^{2}+m_{4}^{2})
)\ 
\:
{\bf J}_{\{2,{m_2}\}\{1,{m_4}\}\{1,{m_3}\}}^{(d)}\ 
\big)
\end{eqnarray}
which is a much shorter expression than the one 
given by eq.~(B.2) in the appendix of \cite{schorsch}.
Moreover, no spurious massless propagator is introduced.

These and further examples can also be found in the
material on the WWW-sites listed in the
program summary.

\end{document}